# Invariants, symmetries and scaling of Rayleigh-Taylor turbulent mixing

Snezhana I. Abarzhi

The University of Chicago, Chicago, IL, USA

Email: snezha@uchicago.edu

We extend the ideas of Kolmogorov theory on symmetries of turbulent dynamics to analyze invariants, scaling and spectra of unsteady turbulent mixing induced by the Rayleigh-Taylor instability. Time- and scale-invariance of the rate of momentum loss leads to non-dissipative momentum transfer between the scales, to $1/2$ and $3/2$ power-law scale-dependencies of the velocity and Reynolds number and spectra distinct from Kolmogorov, and sets the viscous and dissipation scales. Turbulent mixing exhibits more order compared to isotropic turbulence. Mechanisms of flow regularization are discussed.

PACS N: 52.35.Py, 52.35.-g, 47.27.-i, 52.35.Ra, 47.20.-k, 52.57.-z

Rayleigh-Taylor instability (RTI) develops when matters of different densities are accelerated against the density gradients [1,2]. Extensive interfacial material mixing ensues with time [3-5]. This turbulent mixing plays a key role in a broad range of natural phenomena and technological applications spanning astrophysical to atomistic scales and low to high energy densities (see [6,7] and references therein). It influences inertial and magnetic confinement fusion, governs dynamics of supernovae and accretion disks, dominates convection in stellar and planetary interiors, affects aerodynamic turbulent flows as well as flows in atmosphere and ocean. Rayleigh-Taylor (RT) mixing is an anisotropic, non-local, inhomogeneous and statistically unsteady turbulent process [6,7]. Its understanding is required to advance our knowledge of turbulence beyond the limits of idealized canonical approaches [8-12]. Present work considers the influence of momentum transport on the symmetries and invariants, and on the scaling and spectral properties of RT turbulent mixing.

Occurring in a vast variety of physical circumstances, RT flows exhibit certain similarity of evolution [6,7,13]. Initially, small perturbations at the fluid interface grow with time. The growth is exponential if the fluids are incompressible, immiscible and subject to sustained acceleration **g** (gravity) [1]. In the nonlinear regime a coherent structure of bubbles and spikes appears, with light (heavy) fluid with density $\rho_{l(h)}$ penetrating the heavy (light) fluid in bubbles (spikes) [2,13]. The dynamics of the structure is governed by two length scales: the amplitude $\tilde{h}$ in the direction of gravity and the spatial period $\lambda$ in the normal plane [14]. The horizontal scale $\lambda$ is set by the mode of fastest-growth or by the initial conditions [6,13]. It may increase, if the flow is two-dimensional and the initial perturbation is broad-band and incoherent [14]. The vertical scale $\tilde{h}$ grows as power-law with time, and in the mixing regime the growth is believed to be self-similar with $\tilde{h} \sim gt^2$, $|\mathbf{g}| = g$ [3-5,13,15-18]. This scale can be regarded as an integral scale, which represents cumulative contributions of small-scale structures in the flow dynamics. These small-scale vortical structures are produced by shear at the fluid interface [19-21]. In miscible fluids, the small-scale structures diffuse from the interface into the bulk and the mixing process is slowing down. Some other features are induced in the dynamics by compressibility, high energy density conditions and non-uniform acceleration [6,7].

To quantify RT mixing flow the observations were focused on diagnostics of the coarsest scales $\tilde{h}$ and ascertainment of dependency $\tilde{h} = \alpha A g t^2$ where $\alpha$ is a constant and $A = (\rho_h - \rho_l)/(\rho_h + \rho_l)$ is the Atwood number [3-5,15-18,21-23]. RT dynamics was characterized by period $\lambda$, and growth of this scale with $\lambda \sim \tilde{h} \sim gt^2$ was suggested as a primary mechanism of the mixing development [15,18,23]. To account for the time-dependence of the integral scale $\tilde{h} \sim gt^2$ and interpret experimental and numerical data in RTI in terms of turbulent power-laws, some modifications were applied to Kolmogorov theory, including an introduction of a virtual origin and a time-scale for transition to turbulence, a substitution of time-dependencies in Kolmogorov invariants [24-30], and a description of RTI by analogy with passive scalar mixing [31]. Some quantitative agreements were found between the observations data that spanned relatively short dynamic range (about one decade) and the models that used adjustable parameters [4,18,22-30]. Some qualitative features of the turbulent process still remain unclear, e.g. a relatively ordered character of RT flow at high Reynolds numbers [7,32]. To date, experiments and simulations did not provide a trustworthy guidance on whether the concepts of classical turbulence are applicable to an accelerating RT flow and whether the scaling $\tilde{h}/gt^2$ and $\lambda/gt^2$ are indeed universal.

Here we extend to non-canonical circumstances of unsteady turbulent mixing the ideas of Kolmogorov theory [8-12,33] on symmetries of turbulent dynamics. Our consideration accounts for the results of group theory analysis [6,14], that found essentially multi-scale character of RT evolution, and phenomenological modeling [19, 34], that identified the transport of momentum as a better indicator of RT mixing flow than the transport of energy. The invariance of the rate of momentum loss leads to essentially non-Kolmogorov invariant, scaling and spectral properties of the turbulent mixing. The RT mixing exhibits more order compared to isotropic turbulence, and its viscous and dissipation scales are finite and set by the flow acceleration.

(a) Symmetries, invariants and mechanisms:

As in any natural process, turbulent transports are governed by the conservations principles [6,8-11,33,35]. The conservations of mass and momentum have the form

$$\dot{\rho} + \nabla \cdot \rho \mathbf{V} = 0, \qquad \rho(\dot{\mathbf{V}} + (\mathbf{V} \cdot \nabla)\mathbf{V} - \mathbf{g}) + \nabla p + \mathbf{S} = 0, \qquad (1)$$

where $\rho$, $\mathbf{V}$ and $p$ are the fluid density, velocity and pressure, $\mathbf{S}$ denotes terms induced by viscous stress and other effects, and dot marks the partial derivative in time $t$. In RT flow the fluid interface is a discontinuity, and equations (1) yield also the boundary conditions at the interface which balance the transports of mass, momentum and energy of the fluids [6,33]. The system is spatially extended and has no mass sources.

For a homogeneous fluid with $\rho = const$, with neglected effects of gravity, viscous stress and other terms, $\mathbf{g} = \mathbf{S} = 0$, and asymptotically in time, system (1) describes canonical turbulent flow [8-11] that is invariant (in statistical sense) to Galilean transformation, to temporal translations, and to spatial translations, inversions and rotations. The isotropic homogeneous turbulence is also scale-invariant with $L \to LK$, $t \to tK^{1-n}$ and $V \to VK^n$, where $V = |\mathbf{V}|$, $n = 1/3$, and $L$ is a characteristic length scale [8]. The measure of the scaling symmetry is the rate of change of specific kinetic energy $\varepsilon \sim V^3/L$ [8-12]. In contrast to canonical turbulence, the dynamics of RT turbulent mixing is non-inertial. Due to the presence of gravity, $\mathbf{g} \neq 0$, system (1) is invariant with respect to translations, inversions and rotations in the plane normal to $\mathbf{g}$, and exhibits scale-invariance $L \to LK$, $t \to tK^{1-n}$ and $V \to VK^n$ with $n = 1/2$, so that measure $V^2/L$ has the same dimension as $g$ and quantifies the rate of change of specific momentum.

In RT mixing flow, the specific momentum is gained due to buoyancy and is lost due to dissipation. The dynamics of a parcel of fluid is governed by a balance per unit mass of the rate of momentum gain $\tilde{\mu}$ and the rate of momentum loss $\mu$ as

$$\dot{h} = v, \qquad \dot{v} = \tilde{\mu} - \mu \qquad (2).$$

Here $h$ is the vertical length scale, e.g. position of the center of mass of the fluid parcel, $v$ is the corresponding velocity, and $\tilde{\mu}$ and $\mu$ are the absolute values of vectors pointed in opposite directions along the gravity **g** [19,33-35]. Eqs. (2) represent in a simplified dimensional-grounds-based form the conservation of mass and momentum (1). The rate of momentum gain $\tilde{\mu}$ is related to the rate of change of potential energy $\tilde{\varepsilon}$ as $\tilde{\mu} = \tilde{\varepsilon}/v$. The value $\tilde{\mu}$ is the rate of change of momentum which can be gained due to buoyancy, and $\tilde{\mu} = g f(A)$ with $f(A)$ being a function on the Atwood number rescaled hereafter as $g f(A) \to g$. The rate of momentum loss $\mu$ is related to the rate of change of kinetic energy $\varepsilon$ as $\mu = \varepsilon/v$. The value $\mu$ is the rate of change of momentum which is lost due to dissipation. In the limit of vanishing viscosity $\varepsilon = C v^3/L$, where $L$ is the characteristic length scale and $C = const$ [19,33-35].

If the characteristic length scale is horizontal, $L \sim \lambda$, then Eqs. (2) has steady solution with $v \sim \sqrt{g\lambda}$ and $h \sim t\sqrt{g\lambda}$, and rates of momentum and energy are balanced with $\tilde{\mu} = \mu = g$ and $\tilde{\varepsilon} = \varepsilon = g^{3/2}(\lambda/C)^{1/2}$. If the characteristic scale is vertical, $L \sim h$, then asymptotically in time, $h = a g t^2/2$ and $v = a g t$ with $a = (1+2C)^{-1}$. The rates of energy gain and dissipation are time-dependent, $\tilde{\varepsilon} = a g^2 t$ and $\varepsilon = (1-a)a g^2 t$, and the rates of momentum gain and loss are time- and scale-invariant, $\tilde{\mu} = g$ and $\mu = C v^2/h$ [19]. Found in many observations, the values of $a$ are rather small, $a \sim 0.1 - 0.2$, [4,18,21-31]. Thus, in turbulent mixing flow almost all energy induced by the buoyancy dissipates, $\tilde{\varepsilon} \approx \varepsilon$, and the rates of momentum gain and loss slightly unbalance one another, $\tilde{\mu} \approx \mu$ with $(\tilde{\mu} - \mu)/\tilde{\mu} = (1-a)$. Self-similar mixing may develop when horizontal scale $\lambda$ grows with time as $\lambda \sim g t^2$ [15,23], and when the vertical scale $h$ is the characteristic scale for energy dissipation that occur in the small-scale structures at the interface [19,34].

In isotropic turbulence, time- and scale-invariance of the energy dissipation rate implies that the energy injected at large scales, $\varepsilon \sim v^2(v/L)$, is transferred without loss through the inertial range and is dissipated at small scales, $\varepsilon \sim (vL)(v/L)^2$ [8-11]. In turbulent mixing flow the energy dissipation rate is time-dependent, $\varepsilon \sim g^2 t$, whereas the rate of momentum loss is time- and scale-invariant, $\mu \sim v^2/L$. Therefore, at any scale the flow momentum per unit mass is being lost at the same constant rate, and momentum transfer between the scales is non-dissipative. In isotropic turbulence enstrophy is another invariant [8-11], whereas in unsteady turbulent mixing this quantity is time-dependent [33]. It is worth to note that helicity in RT mixing flow approaches a steady value $\sim \mu \sim \tilde{\mu}$ when both vertical and horizontal scales grow with time as $\lambda \sim h \sim g t^2$.

(b) Scaling and spectral properties: To analyze the scaling properties, we presume that at large length scale $L$ and time scale $T$ the characteristic velocity is $v$. At a small length scale $l$ the characteristic velocity is $v_l$, and for a short time-scale $\tau$ the characteristics velocity is $v_\tau$. Time- and scale invariants define how quantities of the turbulent flow scale [33]. In isotropic turbulence [8-11], the invariance of the energy dissipation rate yields the velocity scaling $v_l/v \sim (l/L)^{1/3}$, $n$th order velocity structure function $\sim (l\varepsilon)^{n/3}$, and velocity scaling with time $v_\tau/v \sim (\tau/T)^{1/3}$. Similarly, Reynolds number $\mathrm{Re} = vL/\nu$ and local Reynolds number $\mathrm{Re}_l = v_l l/\nu$ scale as $\mathrm{Re}_l \sim \mathrm{Re}(l/L)^{4/3}$ with the viscous length scale $l_\nu \sim (\nu^3/\varepsilon)^{1/4}$ and time-scale $\tau_\nu \sim (\nu/\varepsilon)^{1/2}$ for $\mathrm{Re}_l \sim 1$.

In unsteady turbulent mixing, the flow quantities scale as defined by the time- and scale-invariance of the rate of momentum loss $\mu \sim v^2/L \sim v_l^2/l$. Particularly, the velocity scales as $v_l/v \sim (l/L)^{1/2}$ and the $n$th order velocity structure function as $\sim (l\mu)^{n/2}$. The velocity scaling with time is $v_\tau/v \sim (\tau/T)$. In RT turbulent mixing the Reynolds number is time-dependent, $\mathrm{Re} = vL/\nu \sim g^2 t^3/\nu$ and the local Reynolds number $\mathrm{Re}_l = v_l l/\nu$ scales as $\mathrm{Re}_l \sim \mathrm{Re}(l/L)^{3/2}$. The length scale $l_\nu$, at which viscous effects are dominant, corresponds to $\mathrm{Re}_l \sim 1$ and is determined by the rate of momentum loss as $l_\nu \sim (\nu^2/\mu)^{1/3}$ with time scale $\tau_\nu \sim (\nu/\mu^2)^{1/3}$. Since $\mu \sim \tilde{\mu} \sim g$, the viscous length scale is set by flow acceleration and is comparable to the mode of fastest growth $\sim (\nu^2/g)^{1/3}$ in RTI [6,13].

Though the Reynolds number of RT flow grows with time, $Re \sim g^2 t^3/\nu$, the velocity scaling indicates that the accelerated mixing with $v_l/v \sim (l/L)^{1/2}$ is more ordered compared to isotropic turbulence with $v_l/v \sim (l/L)^{1/3}$. To further illustrate this property, consider the dynamics of two parcels of fluids involved in the motion with a time delay $\tau$. The particles move relative one another with a constant velocity $\sim g\tau$, comparable to $v_\tau \sim \mu\tau$, whereas their own velocities grow with time as $\sim gt$ and $\sim g(t-\tau)$ and thus reduce the contribution of fluctuations to the mixing transport. This differs from the scenario of isotropic turbulence, where relative velocity of the fluid parcels $\sim (\varepsilon\tau)^{1/2}$ is substantially smaller than the characteristic velocity $v_\tau \sim (\varepsilon\nu\tau)^{1/3}$ [8-11,33].

In isotropic turbulence, the invariance of energy dissipation rate leads to kinetic energy spectrum $E(k) \sim \varepsilon^{2/3} k^{-5/3}$ with $\int_k^\infty E(k)dk \sim \varepsilon^{2/3} k^{-2/3} \sim \varepsilon^{2/3} l^{2/3} \sim v_l^2$ [11]. In RT mixing, dimensional grounds suggests that invariance of the rate of momentum loss $\mu$ leads to $M(k) \sim \mu^{1/2} k^{-3/2}$ for momentum spectrum and $E(k) \sim \mu k^{-2}$ for kinetic energy spectrum with $\int_k^\infty M(k)dk \sim (\mu l)^{1/2} \sim v_l$, $\int_k^\infty E(k)dk \sim \mu l \sim v_l^2$, distinct from Kolmogorov.

(b) Scalar transports:

Transport of scalars (e.g. molecular diffusion) influence significantly the buoyancy-driven turbulent flows. In isotropic turbulence the scalar transport is considered within a context of passive scalar mixing [8-11]. For large Schmidt number $Sc = (\nu/\kappa)$, where $\kappa$ is the scalar diffusivity, the passive scalar dynamics depends on the scales that are acted upon the strain field in the sub-viscous range. The dissipation (Batchelor) length scale $l_B = (\tau_B \kappa)^{1/2}$ is defined as the scale at which the diffusion time-scale $\tau_B$ equals to the viscous time-scale $\tau_\nu$, so that $(l_\nu/l_B)^2 = Sc$ and $l_B = l_\nu Sc^{-1/2}$ [9,10]. Presuming that passive scalar does not influence the momentum transports in RT flow (e.g. the scalar is a neutral buoyancy dye with zero heat solubility) and employing expression for viscous scale $l_\nu \sim (\nu^2/\mu)^{1/3}$, one finds for dissipation scale in turbulent mixing $l_B \sim (\nu \kappa^2/\mu^2)^{1/6}$. Similarly to $l_\nu$, this scale is finite and time-independent, and $l_B \ll l_\nu$ for $Sc \gg 1$.

In realistic mixing flows the transports of scalars influence the transports of mass, momentum, entropy and energy. With account for the heat transfer, the energy conservation (per unit volume) has the form [33]:

$$\partial(\rho s)/\partial t - \chi(\nabla\theta/\theta)^2 - [...] = 0 \qquad (3).$$

Here $\theta$ is temperature, $s$ is entropy, $\chi$ is thermal conductivity, and other terms are omitted in parenthesis [33]. Thermal dissipation is described by function $\varepsilon_\theta = \kappa(\nabla\theta)^2$, and $\varepsilon_\theta \sim (v/L)(\delta\theta)^2$ for turbulent flow with diffusivity $\kappa \sim vL$ [33]. In contrast to problems of isotropic turbulence and turbulent convection [8-11,37,38], in RT turbulent mixing, the thermal dissipation function $\varepsilon_\theta$ is not an invariant, and the rates of momentum and energy are time-dependent.

For an isentropic process in (3), the conditions $\delta\rho/\rho \sim \delta\theta/\theta$ yield for $\tilde{\mu} \sim g\,\delta\rho/\rho$

$$\dot{h} = v, \qquad \dot{v} = \tilde{\mu} - \mu, \qquad \dot{\tilde{\mu}} = -\tilde{C}(v/gh)\tilde{\mu}^2 \qquad (4).$$

where $L \sim h$ and $\tilde{C} > 0$ is a constant [19]. Asymptotically at large times

$$h \sim \frac{gt^2}{\varphi(t)}, \quad v \sim \frac{gt}{\varphi(t)}, \quad \frac{\delta\theta}{\theta_0} \sim \frac{1}{\varphi(t)}, \quad \mu,\tilde{\mu} \sim \frac{g}{\varphi(t)}, \quad \varepsilon,\tilde{\varepsilon} \sim \frac{g^2 t}{\varphi^2(t)}, \quad \varepsilon_\theta \sim \frac{\theta_0^2}{\varphi^2(t)t} \qquad (5)$$

with $\varphi(t) = \ln(gt^2/h_0)$ and $h_0, \theta_0$ being initial length scale and temperature difference. To the leading order, thermal dissipation function $\varepsilon_\theta \to 0$, rates of momentum $\tilde{\mu}, \mu \to 0$ and rates of energy $\tilde{\varepsilon}, \varepsilon \to \infty$. The ratio $\Pi = \mu/\tilde{\mu} = \varepsilon/\tilde{\varepsilon}$ approaches constant value $\Pi \approx 1$. This implies that for turbulent mixing influenced by thermal transport the value of $v^2/(\delta\theta/\theta)gh$ is time and scale-independent. Additional details on scaling and spectral properties of the mixing flow cannot be provided due to unsteadiness of the quantities $\mu, \tilde{\mu}, \varepsilon, \tilde{\varepsilon}, \varepsilon_\theta$. It is worth to note that for time- and space-dependent acceleration $\sim t^a l^b$ ($a, b$ are numbers), the ratio between the rates of energy dissipations and gain $\varepsilon/\tilde{\varepsilon}$ is constant, as in some systems described by the Hamilton-Jacobi equations [35].

(c) Comparison with observations and link to classical approaches:

Invariant measures of turbulent processes provide grounds for methods of their experimental diagnostics [5,12,33]. In RT mixing the traditional turbulent quantities and the integral scale are time-dependent. Time- and scale-invariance of $\mu$ suggests that the rate of momentum can serve as an indicator of the turbulent mixing [19]. Momentum is conjugated with space [35], and to capture momentum transport, spatial distributions of the velocity and density fields should be diagnosed along with their temporal dependencies [6,33]. Measurements of space-time properties of the mixing dynamics require high spatio-temporal resolution and data acquisition rates and substantial dynamic range. These outstanding experimental tasks may likely be addressed in the future [38]. We emphasize that in RT mixing flow dominated by the growth of horizontal scales, $\lambda \sim h \sim gt^2$, the helicity is the other invariant. For existing observations, this quantity may be easier to diagnose than the rate of momentum. The helicity steadiness may serve as an indicator of a merger-driven self-similarity in RTI, analogously to isotropic turbulence, where the flow self-similarity is indicated by the enstrophy saturation [11].

Our momentum-based consideration suggests that RT mixing is more ordered compared to Kolmogorov turbulence. This result agrees with experiments [32], found that at very high Reynolds numbers, $\mathrm{Re} > 10^5$, the mixing flow keeps significant degree of order. It resonates with classic results [39,40] on turbulence relaminarization under high favorable pressure gradient, when the pressure forces dominate the Reynolds stresses, and the flow reverses from turbulent to laminar. This phenomenon is well known for turbulent flows in boundary layers and curved pipes [39,40], which are distinct from isotropic turbulence. To trigger regularization of the turbulent mixing, one may accelerate the flow with a high favorable pressure gradient, similarly to [39,40]. Another possibility is to accurately choose the experimental conditions and ensure the development of accelerated mixing via the dominance of the vertical scales. Such conditions may be achieved with the proper choice of the dispersion relation, of the length-scale and the symmetry of the initial perturbation, and of the outside boundaries [6].

Our results show that the less RT flow is accelerated the more is it likely to resemble the properties of Kolmogorov turbulence. Indeed, turbulent mixing flow is accelerated due to unbalance of momentum and energy, $\mu \neq \tilde{\mu}$ and $\varepsilon \neq \tilde{\varepsilon}$, whereas in nonlinear steady RTI these quantities are balanced, so that the total momentum rate is zero $\tilde{\mu} - \mu = 0$ and $\varepsilon = const$, similarly to Kolmogorov turbulence [8-11]. This explains the recent observations [4,5,20,21,24-37] of statistically steady turbulence in RT flows with relatively low Reynolds numbers $\mathrm{Re} \sim 10^3$. Distinct from $k^{-5/3}$ kinetic energy spectrum found in these observations indicates the effect of slight momentum unbalance on the flow dynamics, as found in the foregoing.

Momentum-based consideration of RT turbulent mixing treats self-consistently the time-dependence of the integral scale $\sim gt^2$, finds stronger coupling of the large and small scales than in isotropic turbulence, and shows that the viscous and dissipation scales of the mixing flow are finite and set by the flow acceleration. The range of length scales and time scales in RT turbulent mixing span as $(L/l_v) \sim \text{Re}^{2/3}$ and $(T/\tau_v) \sim \text{Re}^{1/3}$. For fluids with contrasting densities the span of scales agrees with observations [3-5,17-19, 21-31]. An upper limit for Reynolds number $\text{Re} \sim g^2 t^3/\nu$ can be estimated at a border of validity of incompressible approximation as $\text{Re}_c \sim c^3/g\nu$, where $c$ is the sound speed.

We have extended to non-canonical circumstances the ideas of Kolmogorov theory on symmetries of turbulent dynamics [8-12,33], and have analyzed the influence of momentum transport on the properties of anisotropic, inhomogeneous and non-inertial turbulent mixing induced by the RTI. The invariants, scaling and spectra of the turbulent mixing are found to depart from the Kolmogorov scenario. The flow invariant is the rate of momentum loss, and it plays the same role in the anisotropic turbulent process as the rate of energy dissipation in isotropic turbulence. Unsteady turbulent mixing exhibits more order compared to Kolmogorov turbulence, and its viscous and dissipation scales are finite and set by the flow acceleration. The results obtained can serve for development of rigorous mathematical approaches and systematic experimental and numerical studies of non-equilibrium turbulent processes.

## References

1. Rayleigh, Lord, 1883, *Proc. London Math. Soc.* **14**, 170.
2. Davies R. M. and Taylor G. I., 1950, *Proc. R. Soc. London*, Ser. **A 200**, 375.
3. Read, K. I., 1984, *Physica D* **12**, 45.
4. Ramaprabhu, P., Andrews, M. J., 2004, *J. Fluid Mech.* **502**, 233.
5. Meshkov, E. E., 2006, *Studies of hydrodynamic instabilities*, FGYC-VNIIEF. Sarov (in Russian).
6. Abarzhi, S. I., 2008, *Physica Scripta* **T132,** 297681.
7. Remington B. A., Drake R.P., Ryutov D.D., 2006, *Rev. Mod. Phys.* **78**, 755.
8. Kolmogorov, A.N., 1941, *Dokl. Akad. Nauk. SSSR* **30**, 299 and **32**, 19.
9. Batchelor, G.K., 1953, *The theory of homogeneous turbulence*, Cambridge: Cambridge Univ. Press.
10. Monin, A. S., Yaglom, A. M., 1979, *Statistical fluid mechanics*, MIT Press, Cambridge.
11. Frisch, U., 1995, *Turbulence, the legacy of A.N. Kolmogorov*, Cambridge University Press, Cambridge.
12. Sreenivasan, K. R., 1999, *Rev. Mod. Phys.* **71**, S383.
13. Youngs, D. L., 1984, *Physica* D **12,** 32.
14. Abarzhi, S. I., Nishihara, K., Rosner, R., 2006, *Phys. Rev.* E **73,** 036310.
15. Alon, U., Hecht J., Offer D., Shvarts D., 1995, *Phys. Rev. Lett.* **74,** 534.
16. Fermi, E., von Newman, J., 1951, In Fermi, E., 1962, *Collected papers* **2**, 816.
17. Dalziel, S. B., Linden, P. F., Youngs, D. L., 1999, *J. Fluid Mech.* **399**, 1.
18. Dimonte, G., Youngs, D. L., et al. 2004, *Phys. Fluids* **16**, 1668.
19. Abarzhi, S. I., Gorobets, A., Sreenivasan, K. R., 2005, *Phys. Fluids* **17**, 081705.
20. He, X. Y., Zhang, R. Y., Chen, S. Y., Doolen, G. D., 1999, *Phys. Fluids* **11,** 1143.
21. Kadau, K., Rosenblatt, C., Barber, et al., 2007, *Proc. Natl. Acad. Sci. USA* **104,** 774107745.
22. Calder, A. C., *et al.,* 2002, *Astrophys. J., Suppl. Ser.* **143,** 201.
23. Glimm, J., Sharp, D. H., 1990, *Phys. Rev. Lett.* **64,** 2137.
24. Steinkamp, M. J., Clark, T. T., Harlow F. H., 1999, *Int. J. Multiph. Flow* **25,** 599 and **25,** 639.
25. Ristorcelli, J. R., Clark, T. T., 2004, *J. Fluid Mech.* **507,** 213.
26. Dimotakis, P. E., 2000, *J. Fluid Mech.* **409,** 69.
27. Cabot, W.H., Cook, A. W., 2006, *Nature Physics,* **2**, 562.
28. Chertkov, M., 2003, *Phys. Rev. Lett.* **91,** 115001.
29. Boffetta, G., Mazzino, A., Musacchio, S., Vozella, L. 2009, *Phys. Rev. E* **79**, 065301.
30. Poujade, O., 2006, *Phys. Rev. Lett.* **97**, 185002.
31. Gauthier, S., Bonnet, M., 1990, *Phys. Fluids* A **2** 1685.
32. Robey, H. F., et al., 2003, *Physics of Plasmas* **10**, 614.
33. Landau, L. D., Lifshitz, E. M., 1987, *Course Theor. Phys. VI Fluid Mech*, Pergamon Press, NY.
34. Abarzhi, S. I., Cadjun, M., Fedotov, S., 2007, *Phys. Lett. A* **371,** 457.
35. Landau, L. D., Lifshitz, E. M., 1987, *Course Theor. Phys. I, Mechanics*, Pergamon Press, NY.
36. Bolgiano, R. Jr., 1959, *J. Geophysical Research* **64**, 2226.
37. Yakhot, V. 1992, *Phys Rev. Lett.* **69**, 769.
38. Orlov, S.S., 2008, *Physica Scripta* **T132**, 014056.
39. Taylor, G. I., 1929, *Proc. Roy. Soc. A* **124**, 243.
40. Narasimha, R., Sreenivasan, K. R., 1973, *J. Fluid Mechanics*, **61**, 417.